# Chromium-Induced Ferromagnetism with Perpendicular Anisotropy in Topological Crystalline Insulator SnTe (111) Thin Films

Fei Wang[1,2,4], Hongrui Zhang[3,4], Jue Jiang[2], Yi-Fan Zhao[2], Jia Yu[2], Wei Liu[1*], Da Li[1], Moses H. W. Chan[2], Jirong Sun[3], Zhidong Zhang[1], and Cui-Zu Chang[2*]

[1]Shenyang National Laboratory for Materials Science, Institute of Metal Research, Chinese Academy of Sciences, Shenyang 110016, China

[2]Department of Physics, Pennsylvania State University, University Park, PA 16802, USA

[3]Beijing National Laboratory for Condensed Matter, Institute of Physics, Chinese Academy of Sciences, Beijing 100190, China

[4]University of Chinese Academy of Sciences, Beijing 100049, China

Corresponding authors: wliu@imr.ac.cn (W. L.) and cxc955@psu.edu (C. Z. C.).

**Topological crystalline insulator (TCI) is a recently-discovered topological phase of matter. It possesses multiple Dirac surface states, which are protected by the crystal symmetry. This is in contrast to the time reversal symmetry that is operative in the well-known topological insulators. In the presence of a Zeeman field and/or strain, the multiple Dirac surface states are gapped. The high-Chern-number quantum anomalous Hall (QAH) state is predicted to emerge if the chemical potential resides in all the Zeeman gaps. Here, we use molecular beam epitaxy to grow 12 double layer (DL) pure and Cr-doped SnTe (111) thin film on heat-treated SrTiO$_3$ (111) substrate using a quintuple layer of insulating**

**(Bi$_{0.2}$Sb$_{0.8}$)$_2$Te$_3$ topological insulator as a buffer film. The Hall traces of Cr-doped SnTe film at low temperatures display square hysteresis loops indicating long-range ferromagnetic order with perpendicular anisotropy. The Curie temperature of the 12DL Sn$_{0.9}$Cr$_{0.1}$Te film is ~ 110 K. Due to the chemical potential crossing the bulk valence bands, the anomalous Hall resistance of 12DL Sn$_{0.9}$Cr$_{0.1}$Te film is substantially lower than the predicted quantized value (~1/4 $h/e^2$). It is possible that with systematic tuning the chemical potential via chemical doping and electrical gating, the high-Chern-number QAH state can be realized in the Cr-doped SnTe (111) thin film.**

The quantum anomalous Hall (QAH) state is a topological quantum state with quantized Hall resistance and zero longitudinal resistance under a zero magnetic field [1-3]. The QAH dissipation-free chiral edge state enables potential applications in the next-generation electronic and spintronic devices with low-power consumption [4]. Although the possibility of QAH effect was first envisioned by Haldane in 1988 [1], little experimental progress was made in the following two decades. This is the case because the realization of the QAH state in a real material requires a ferromagnetic insulator with inverted band structure [4-6]. This stringent condition is found to be satisfied in magnetically doped topological insulators (TI) [2,7,8]. Indeed, the QAH effect was observed in 5 quintuple layers (QL) Cr-doped (Bi,Sb)$_2$Te$_3$ thin films in 2013 [3]. Since this initial discovery of the QAH state [3], intensive theoretical and experimental research has been focusing on finding the QAH state in other material systems [4-6], specifically high-Chern-number QAH state in Cr-doped Bi$_2$(Se,Te)$_3$ [9] and magnetically doped topological crystalline insulator (TCI) [10]. The high-Chern-number QAH effect in Cr-doped $Bi_2(Se,Te)_3$ is the consequence of the stronger exchange field induced two or more pairs of inverted subbands [9], whereas the high-Chern-number QAH effect in magnetically doped TCI is due to the existence of the multiple Dirac surface states [10]. In the former system, the non-square or hysteresis-free loop makes the realization of the high-Chern-number QAH effect nearly impossible [11]. Moreover, the required stronger exchange field will more likely drive the magnetic TI into a semimetal instead of the high-Chern-number QAH phase [9]. In the latter system, since the multiple Dirac surface states are the intrinsic property of TCI [12], it would seem feasible to realize the high-Chern-number QAH effect provided that the ferromagnetism with perpendicular anisotropy is found in the magnetically doped TCI films [10]. Compared with the single chiral edge channel of the QAH effect realized in magnetically doped 3D TI films [3,13-15], the presence of the multiple dissipation-free chiral edge channels in the high-Chern-number QAH state would lower the contact resistance and increase the breakdown current of the edge conduction. Both properties are highly desirable for improving the performance of the future QAH state-based chiral interconnects [16].

The topological nature of TCI is protected by the crystal symmetry [12] instead of the time reversal symmetry. The fact that it harbors even number of Dirac surface states leads to many possible fascinating topological phenomena in TCI, *e.g.* spin-filtered edge states [17], quantum spin Hall effect [18,19] and high-Chern-number QAH effect [10]. SnTe, with the rock-salt structure (**Fig. 1a**), has been predicted to be a TCI with even number nontrivial Dirac surface states on high-symmetry crystal surfaces, such as (001), (110) and (111) [12]. The Dirac surface states on SnTe (001) and (111) surfaces have been observed in angle-resolved photoemission spectroscopy (ARPES) measurements [20-22]. Unlike the Dirac cones on the (001) surface that are centered at the non-time-reversal-invariant-momenta (TRIM) points, all four Dirac cones on the (111) surface of SnTe are located at the TRIM ($\bar{\Gamma}$ and $\bar{M}$) points, respectively, as shown in **Fig. 1b**. Therefore, the Dirac surface states on SnTe (111) surface are protected not only by the crystal symmetry but also by the time reversal symmetry [23]. In the presence of a Zeeman field (*i.e.* ferromagnetic order) and/or strain, the four Dirac cones on SnTe (111) surface are gapped, as shown in **Fig. 1c**. If the chemical potential is tuned into all the Zeeman gaps at $\bar{\Gamma}$ and $\bar{M}$ points simultaneously, the QAH state with Chern-number $C=\pm 4$ will be realized.

The Dirac point at $\bar{M}$ point is predicted to be higher (Te-terminated) or lower (Sn-terminated) in binding energy than the Dirac point at $\bar{\Gamma}$ point in SnTe (111) surface [23,24]. This binding energy difference is indeed observed, specifically, ~170 meV in bulk SnTe (111) sample by ARPES measurements [22] and ~140 meV in 30 nm film grown on 30 nm $Bi_2Te_3$ (111) by transport measurements [25]. Since the Zeeman gap is usually several meVs, depending on the $T_C$ of the ferromagnetic order, it is impossible to achieve the high-Chern-number QAH effect in the bulk or thick SnTe (111) samples. However, quantum confinement effect can modify the bulk band structure of thinner film [26], possibly making the Dirac points at $\bar{\Gamma}$ and $\bar{M}$ points located at the same energy level. In addition, we note that in $Pb_{1-x}Sn_xSe$ (111) thin films, ARPES measurements reveal the Dirac points at $\bar{\Gamma}$ and $\bar{M}$ points has no relative binding energy difference [27]. Therefore, the Dirac cones at $\bar{\Gamma}$ and $\bar{M}$ points in SnTe (111) film could be well tuned at the same binding energy level by tuning the thickness and/or compositions of the films.

The SnTe films have been grown by molecular beam epitaxy (MBE) technique on various substrates, *e. g.* Si (111) [28], Si (100) [29], SrTiO$_3$ (001) [30], BaF$_2$ (111) [31,32], BaF$_2$ (001) [33]. Due to lattice mismatch, SnTe films grown on the above substrates have relatively low quality as compared with the MBE-grown TI films [3,34]. Recently, SnTe (111) films grown on the Bi$_2$Te$_3$ (111) buffer layer are found to be of much higher quality [21,25,35]. However, the high electrical conduction of thick Bi$_2$Te$_3$ buffer film (~ 7 nm in [21], ~ 30 nm in [25] and ~ 4 nm in [35]) will inevitably affect the transport property of the SnTe (111) films [25]. In addition to the difficulty of fabricating the SnTe (111) thin films, it is also challenging to dope the transition metal ions into SnTe to achieve well-defined ferromagnetic order with perpendicular anisotropy [36,37]. Indeed, ferromagnetic order with square Hall hysteresis loop in magnetically doped SnTe film is to date lacking.

In this paper, we report experimental studies of pure and Cr-doped SnTe (111) films with thickness of 12 double layers (DL) (~ 4.3 nm) grown on the heat-treated SrTiO$_3$ (111) substrate using a QL insulating (Bi$_{0.2}$Sb$_{0.8}$)$_2$Te$_3$ film as a buffer layer. The SnTe (111) films with around 12 DL thicknesses have been predicted to be quantum spin Hall insulator (*i.e.* 2D TI) [18,19], from which the QAH state is likely reached by killing off one spin channel of the helical edge states [2]. Both 12 DL pure and Cr-doped SnTe (111) films show metallic behavior, indicating the continuous films have been formed. The Hall trace of the Cr-doped SnTe (111) film displays a square hysteresis loop, indicating well-defined ferromagnetic order with perpendicular anisotropy. The anomalous Hall resistance is only several Ωs at $T$=2K and the Curie temperature ($T_C$) is ~110K with doping level $x$=0.1 ($Sn_{1-x}Cr_xTe$).

Pure and Cr-doped SnTe (111) films and the 1 QL insulating $(Bi_{0.2}Sb_{0.8})_2Te_3$ buffer film were grown in a commercial MBE chamber with a base pressure ~ $5 \times 10^{-10}$ mbar. The heat-treated insulating $SrTiO_3$ (111) substrates were outgassed at 600°C for 1 hour before the growth of TCI films. High purity Sn (99.999%), Cr (99.995%), Bi (99.9999%), Sb (99.9999%) and Te (99.9999%) were evaporated from standard Knudsen cells. During the TCI film growth, the substrate was maintained at ~250°C. The flux ratio of Te/Sn was set to be ~ 4 and the TCI growth rate is ~ 0.6 DL/min. The Sn and Cr concentrations in the Cr-doped SnTe films were first determined by their ratio obtained *in-situ* during growth using separate quartz crystal monitors and later confirmed by inductively coupled plasma atomic emission spectroscopy (ICP-AES) measurements. To avoid possible contamination, a ~ 10 nm thick Te layer is deposited at room temperature on top of the SnTe (111) film prior to its removal from the MBE chamber for *ex-situ* transport measurements carried out in a commercial Physical Property Measurement System (PPMS) cryostat (Quantum Design, 2 K, 14 T).

**Figures 2a** to **2c** display the reflection high-energy electron diffraction (RHEED) patterns of heat-treated bare insulating SrTiO$_3$ (111) substrates, 1QL (Bi$_{0.2}$Sb$_{0.8}$)$_2$Te$_3$ film and 12DL Sn$_{0.9}$Cr$_{0.1}$Te film. The clear reconstruction of the heat-treated SrTiO$_3$ (111) substrate (**Fig. 2a**) indicates its atomic flat surface, which is suitable for the TCI film growth. The sharp and streaky '1×1' patterns of 12DL Sn$_{0.9}$Cr$_{0.1}$Te film (**Fig. 2c**) grown on 1QL (Bi$_{0.2}$Sb$_{0.8}$)$_2$Te$_3$ buffer film indicate the highly-ordered and smooth surface of the Cr-doped SnTe film. The good crystal quality of the Cr-doped SnTe (111) film is further demonstrated by X-ray diffraction (XRD) measurements. **Fig. 2d** shows the XRD image of the 48DL Sn$_{0.9}$Cr$_{0.1}$Te film grown on SrTiO$_3$ (111) substrate. The (222) reflection of SnTe film is seen, thus confirming the single crystal (111) phase of the Sn$_{0.9}$Cr$_{0.1}$Te film. A couple of refection peaks of 1QL (Bi$_{0.2}$Sb$_{0.8}$)$_2$Te$_3$ buffer film are also seen in the XRD spectrum. In addition, we also carried out XRD $\phi$-scan measurements on a 12DL Sn$_{0.9}$Cr$_{0.1}$Te film. The XRD $\phi$-scan result (**Fig. 2e**) shows a peak repetition of 60$^o$, indicating the twin domains are formed in our Sn$_{0.9}$Cr$_{0.1}$Te (111) film. Previous studies have demonstrated the (Bi,Sb)$_2$Te$_3$ film grown on SrTiO$_3$(111) substrate has the twin domain structures [38,39], so the two domains found in our Sn$_{0.9}$Cr$_{0.1}$Te (111) film is very likely due to the buffer film of 1QL (Bi$_{0.2}$Sb$_{0.8}$)$_2$Te$_3$. The QAH state has been successfully realized in the Cr- and V-doped (Bi,Sb)$_2$Te$_3$ film grown on the SrTiO$_3$(111) substrates [3,15], so it is very likely that the twin domains structures in our Cr-doped SnTe (111) will not affect the realization of the high-Chern-number QAH state.

In order to characterize the suface morphology of 12 DL $Sn_{0.9}Cr_{0.1}Te$ (111) film, we carried out X-ray reflectivity (XRR) measurements and found the average roughness of the film is ~0.33nm (See Supporting Materials). Such a roughness is comparable to the average roughness of the Cr-doped TI films of the same thickness [40]. In addition, the high resolution scanning transmission electron microscopy (STEM) is also employed to characterize the Cr-doped SnTe film. The highly-ordered lattice structure is seen (**Fig. 2f**) and the uniform Cr distribution inside the host SnTe matrix is confirmed by the energy-dispersive X-ray spectroscopy (EDS) mapping (**Fig. 2g**).

**Figure 3a** shows the temperature ($T$) dependence of the sheet longitudinal resistance ($\rho_{xx}$) of a 12DL undoped SnTe (111) film. The $\rho_{xx}$ decreases with decreasing temperature, indicating its metallic behavior. At sufficient low temperatures ($T < 5$ K) , the $\rho_{xx}$ reaches a minimal value and then increases with decreasing temperature, suggesting the presence of an insulating ground state (the inset of **Fig. 3a**). Similar to the TI thin films, the insulating ground state in TCI films is possibly due to the combination of electron-electron interaction and topological delocalization [41,42]. **Figure 3b** shows the normalized magnetoresistance $MR=\rho_{xx}(H)/\rho_{xx}(0)$ measured at different angles θ between the magnetic field ($\mu_0H$) and the normal of the film plane. When θ = 0°, namely when $\mu_0H$ is perpendicular to the film plane, the *MR* shows a downward cusp near zero magnetic field. This cusp feature is consistent with weak anti-localization effect [25,31]. As θ is gradually increased to 90 °, the cusp at zero magnetic field broadens, indicating the weakening of the contribution from 2D massless Dirac fermions in TCI films [31].

The Hall trace of 12DL SnTe (111) film measured at 2 K is shown in **Fig. 3c**. The positive slope of the linear Hall trace indicates our SnTe (111) film is dominated by *p*-type carriers, this is different from the nonlinear Hall trace in 30 nm SnTe (111) grown on 30 QL Bi$_2$Te$_3$ films [25]. The calculated carrier density and carrier mobility are $n_{2D}$ ~ 1.022 $\times 10^{15}$ cm$^{-2}$ and $\mu$ ~ 29 cm$^2$V$^{-1}$S$^{-1}$. The high $n_{2D}$ and low $\mu$ are due to the additional Sn vacancies in the SnTe (111) films [43]. The $n_{2D}$ and $\mu$ of 12DL SnTe (111) film increase with decreasing the temperature (**Fig. 3d**), and a small decrease is observed at low temperatures, which is possibly due to the presence of the insulating ground state.

In order to realize the QAH state in TCI, ferromagnetic order must be introduced into SnTe (111) films. Unlike the magnetic proximity effect, ferromagnetic order induced by direct doping of transition metal ions is more uniform, thus the four Dirac cones on top and bottom surfaces of TCI films can be gapped simultaneously. As noted above, the *C*=1 QAH state to date has been realized only in Cr- and V-doped (Bi,Sb)$_2$Te$_3$ films [3,15]. Therefore, we dope traces of Cr ions in the SnTe (111) films. **Figure 4a** shows the temperature dependence of the $\rho_{xx}$ of 12DL Sn$_{0.9}$Cr$_{0.1}$Te film. At high temperature, the $\rho_{xx}$ shows metallic behavior, for $T < 30$ K, an insulating behavior similar to that found in the Cr-doped TI films [44,45] is observed. The $\rho_{xx}$ does not saturate at low temperatures but increases much faster than logarithmic function (the inset of **Fig. 4a**), suggesting that the insulating behavior cannot be attributed to Kondo effect [44].

**Figures 4b** and **4c** display the $\mu_0H$ dependence of the $\rho_{xx}$ and $\rho_{yx}$, respectively. At $T = 2$ K, the normalized *MR* of 12DL $Sn_{0.9}Cr_{0.1}Te$ film has a butterfly feature and the $\rho_{yx}$ shows a 'square-like' hysteresis loop, indicating the long-range ferromagnetism with perpendicular anisotropy in 12DL $Sn_{0.9}Cr_{0.1}Te$ film. The observation of the 'square-like' AH effect further confirms the Cr distribution inside our Cr-doped SnTe films is uniform. The hysteresis loop observed in 12DL $Sn_{0.9}Cr_{0.1}Te$ film is comparable to that of the Cr-doped $Sb_2Te_3$ film, which is the "parent" TI material for the realization of the QAH effect [3,45,46]. The coercive field ($H_c$) of 12DL $Sn_{0.9}Cr_{0.1}Te$ film is ~0.155T at $T = 2$ K. Both the anomalous Hall resistance ($\rho_{AH}$) (which we define as the intercept of the Hall trace when 0.25 T $< \mu_0H <$ 0.5 T) and the $H_c$ decrease with increasing temperature, suggesting that the ferromagnetic order is weakened by thermally activated spin orientation fluctuations. The temperature dependence of the $\rho_{AH}$ of the 12DL $Sn_{0.9}Cr_{0.1}Te$ film are summarized in **Fig. 5a**, the nonzero $\rho_{AH}$ at $T = 100$ K and the zero $\rho_{AH}$ at $T = 120$ K indicate the $T_C$ of 12DL $Sn_{0.9}Cr_{0.1}Te$ film is between 100 K and 120 K.

In order to determine more accurately $T_C$, we symmetrically make Hall measurements at 100 K $< T <$ 120 K. The inset of **Fig. 5a** displays the Arrott plot of the 12DL $Sn_{0.9}Cr_{0.1}Te$ film. The Arrott plots at low temperatures are straight lines with positive intercept, indicating the ferromagnetic phase. The intercept changes from positive to negative when $T$ is between 105 K and 115 K, indicating a $T_C$ of 110K for this 12DL $Sn_{0.9}Cr_{0.1}Te$ film. The Curie temperature of Cr-doped SnTe film is two times higher than that of Cr-doped TI with the same doping level [15,45]. This observation excludes the interpretation that the ferromagnetism observed here is due to the Cr interdiffusion into the 1QL $(Bi,Sb)_2Te_3$ buffer film. Moreover, ferromagnetism is formed in the Cr-doped $(Bi,Sb)_2Te_3$ only when the thickness is above 3QL [47].

To realize the high-Chern-number QAH state in Cr-doped SnTe (111) film, the third and also the last requirement is to tune the chemical potential to be located within all four Zeeman gaps (**Fig. 1c**). The carriers in 12DL $Sn_{0.9}Cr_{0.1}Te$ film is also *p*-type and the carrier density $n_{2D} \sim 5.3 \times 10^{14}$ cm$^{-2}$ at $T = 2$ K. Compared with the undoped SnTe film, the Cr-doped SnTe film has lower carrier density, possibly due to the coexistence of $Cr^{2+}$ and $Cr^{3+}$ ions [48,49]. Nevertheless, the chemical potential of Cr-doped SnTe film still resides in the bulk valence bands and the ferromagnetic order in the Cr-doped SnTe film should be mediated by bulk carriers (*i.e.* Ruderman-Kittel-Kasuya-Yosida (RKKY) interaction). With the chemical potential approaching the surface states, other ferromagnetic mechanisms in the intermediate carrier density range such as Blombergen-Rowland interaction may also play a role in Cr-doped SnTe samples [50]. When the chemical potential of the Cr-doped SnTe film is tuned to cross the surface states at $\bar{\Gamma}$ and $\bar{M}$ points, the surface carriers-induced van Vleck interaction will take effect and dominate the ferromagnetic order of the Cr-doped SnTe films [51]. To neutralize the *p*-type carriers in the Cr-doped SnTe films, we need to introduce electrons by chemical doping and/or electrical gating. PbTe has the same structure with SnTe, but with the electron carriers [21], so the chemical potential of the film can be tuned by doping trace of Pb in Cr-doped SnTe [21]. Alternatively, the Bi doping can also work for tuning the chemical potential of Cr-doped SnTe films [52,53]. In addition, as noted above, the binding energy difference of the four Dirac points at $\bar{\Gamma}$ and $\bar{M}$ points can be reduced by tuning the thickness [26] and/or compositions [27] of the films, as shown in **Fig. 5b**. Chemical surface modification is another backup option in tuning the chemical potential. A theoretical calculation proposed that the chemisorption, specifically the bromine or the iodine absorbed on SnTe (111) surface, can eliminate the binding energy difference of the Dirac points at $\bar{\Gamma}$ and $\bar{M}$ points [54].

In summary, we succeeded in fabricating high quality pure and Cr-doped SnTe (111) thin films on SrTiO$_3$ (111) substrates using 1 QL insulating TI (Bi$_{0.22}$Sb$_{0.78}$)$_2$Te$_3$ buffer film by MBE. The well-defined ferromagnetism with perpendicular anisotropy and higher $T_C$ is demonstrated in Cr-doped SnTe (111) thin films. We expect our work will motivate more studies on the fascinating property of the TCI thin films and pave the avenue for realizing the high-Chern-number QAH state in magnetically-doped TCI system.

**Acknowledgement**

The authors would like to thank J. Liu and C. Liu for the helpful discussions, P. Wei for the help with data analysis and T. Pillsbury and W. Zhao for the help with experiments. F. W., H. Z., W. L., D. L. J. S. and Z. Z. acknowledge the support from National Natural Science Foundation of China (NSFC) (Grants No. 51590883, 51331006 and 51771198), the Key Research Program of Chinese Academy of Sciences (Grant No. KJZD-EW-M05-3), and State Key Program of Research and Development of China (Grant No. 2017YFA0206302). C. Z. C. thanks the startup grant supported by Penn State University.

**Figures and figure captions:**

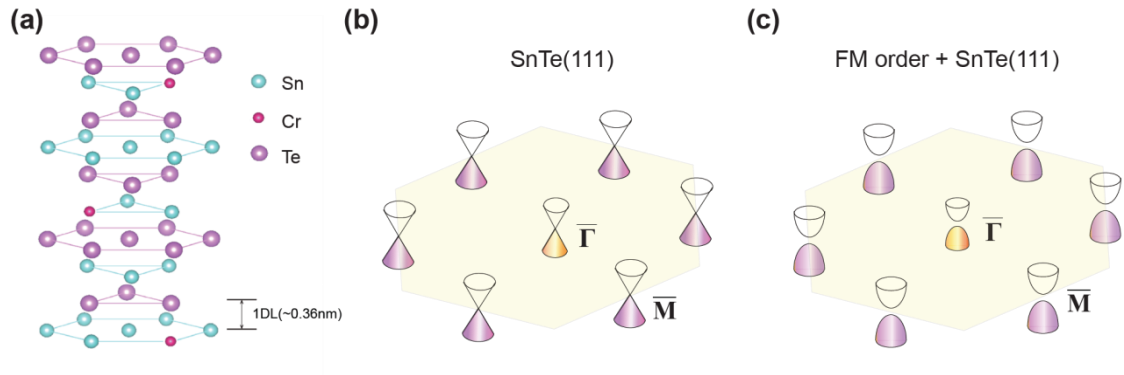

**Fig. 1| Cr-doped SnTe (111) films.** (a) Schematics of the crystal structure of Cr-doped SnTe (111) films. (b) Brillion zone of pure SnTe (111) film when the chemical potential is near the charge neutral point. Four Dirac cones are found at time-reversal-invariant momenta of the Brillion zone, one is at $\bar{\Gamma}$ point, and the other three are at $\bar{M}$ points. (c) Brillion zone of Cr-doped SnTe (111) film when the chemical potential is near the charge neutral point. All Dirac cones are gapped due to the presence of the ferromagnetic order.

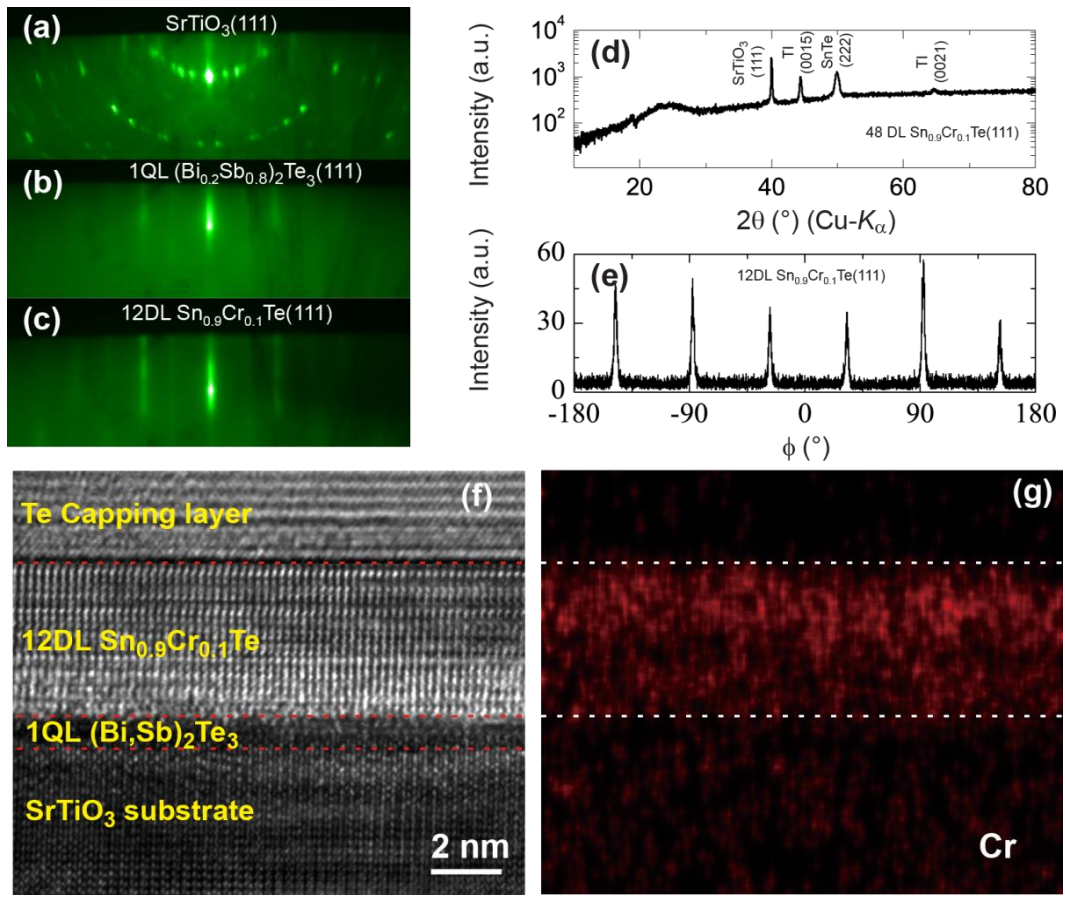

**Fig. 2| Characterizations of MBE-grown Cr-doped SnTe (111) films.** (a-c) RHEED patterns of the heat-treated SrTiO$_3$ (111) substrate (a), 1QL insulating (Bi$_{0.2}$Sb$_{0.8}$)$_2$Te$_3$ (111) buffer film (b) and 12DL Sn$_{0.9}$Cr$_{0.1}$Te (111) film (c). (d) The XRD image of 48DL Sn$_{0.9}$Cr$_{0.1}$Te (111) film. (e) The φ-scan result of 12DL Sn$_{0.9}$Cr$_{0.1}$Te (111) film. (f) The high-resolution cross-sectional STEM image of 12DL Sn$_{0.9}$Cr$_{0.1}$Te (111) film. (g) The corresponding EDS mapping of the Cr element in 12DL Sn$_{0.9}$Cr$_{0.1}$Te (111) film.

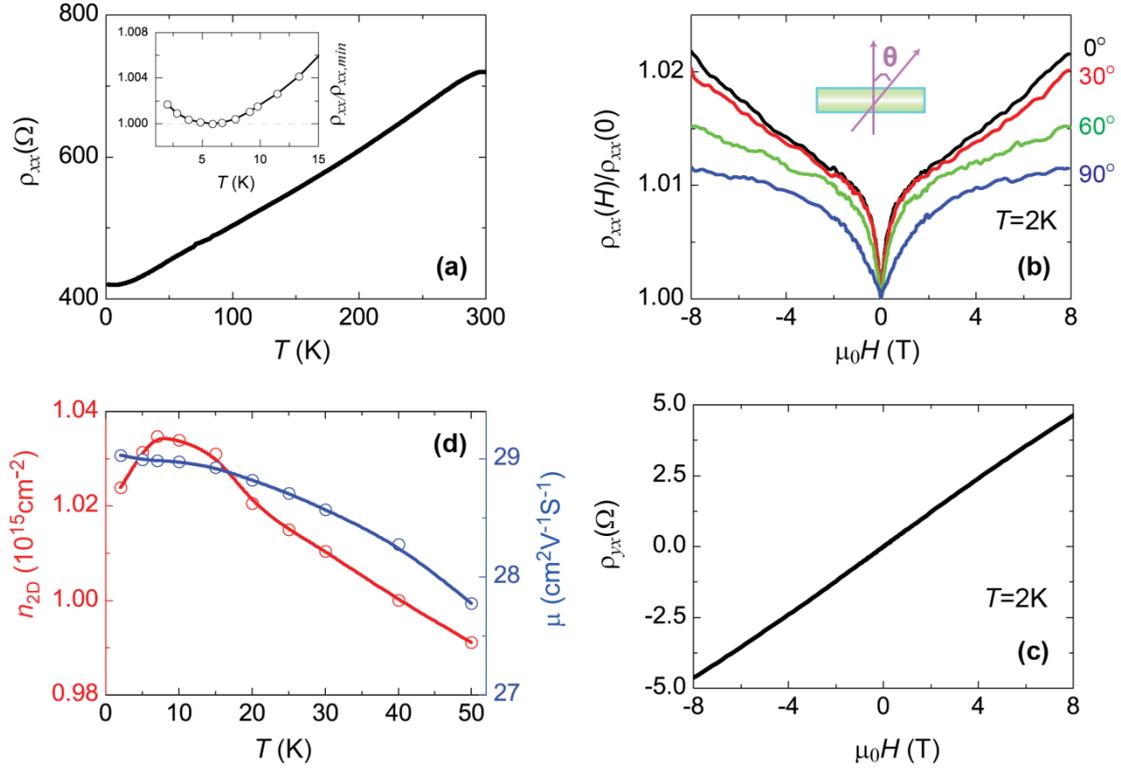

**Fig. 3| Transport properties of 12 DL undoped SnTe (111) film.** (a) Temperature dependence of the sheet longitudinal resistance ($\rho_{xx}$) of 12DL SnTe (111) film. Inset: the normalized $\rho_{xx}/\rho_{xx,min}$ shows an upturn with decreasing $T$ when $T < 5$ K. (b) Angle dependence of the normalized magnetoresistance $MR=\rho_{xx}(H)/\rho_{xx}(0)$ of 12DL SnTe(111) film measured at $T = 2$ K. (c) Hall trace of 12DL SnTe (111) film measured at $T = 2$ K. (d) Temperature dependence of carrier density $n_{2D}$ (red) and carrier mobility $\mu$ (blue) of 12DL SnTe (111) film.

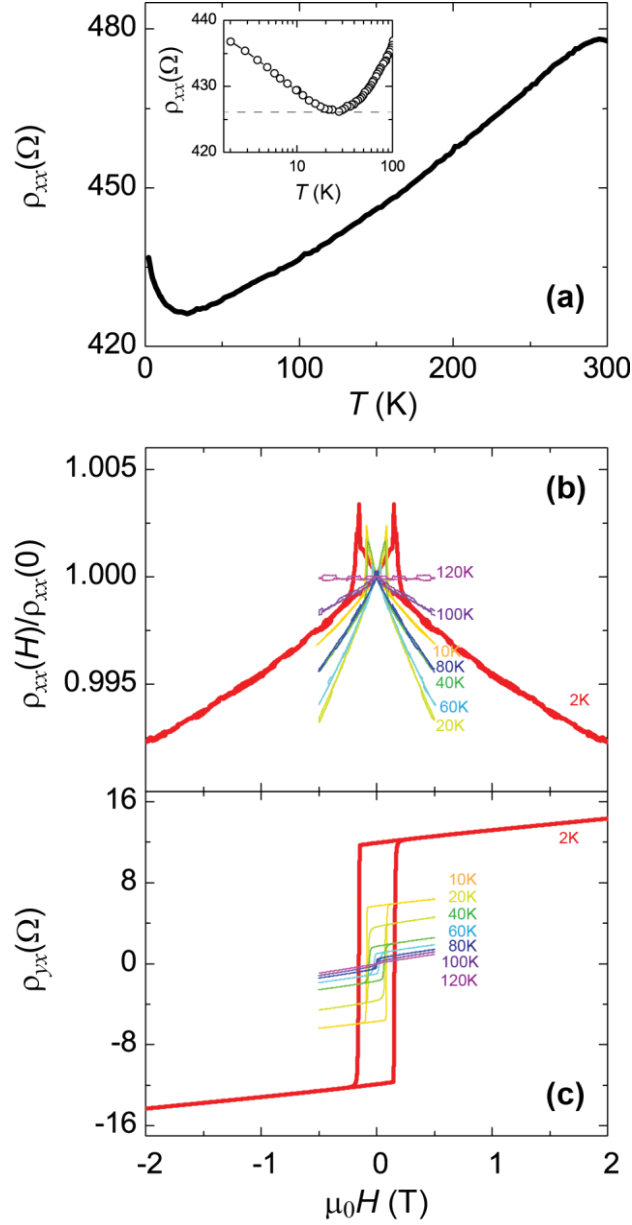

**Fig. 4| Transport properties of 12DL Cr-doped SnTe (111) film.** (a) Temperature dependence of the $\rho_{xx}$ of 12DL $Sn_{0.9}Cr_{0.1}Te$ (111) film. Inset: the temperature dependence of $\rho_{xx}$ of 12DL $Sn_{0.9}Cr_{0.1}Te$ (111) film plotted in a semi-log scale. $\rho_{xx}$ diverges much faster than $\log T$ at low temperatures. (b, c) The $\mu_0 H$ dependence of the normalized $\rho_{xx}$ and the $\rho_{yx}$ of the 12DL $Sn_{0.9}Cr_{0.1}Te$ (111) film measured at different temperatures.

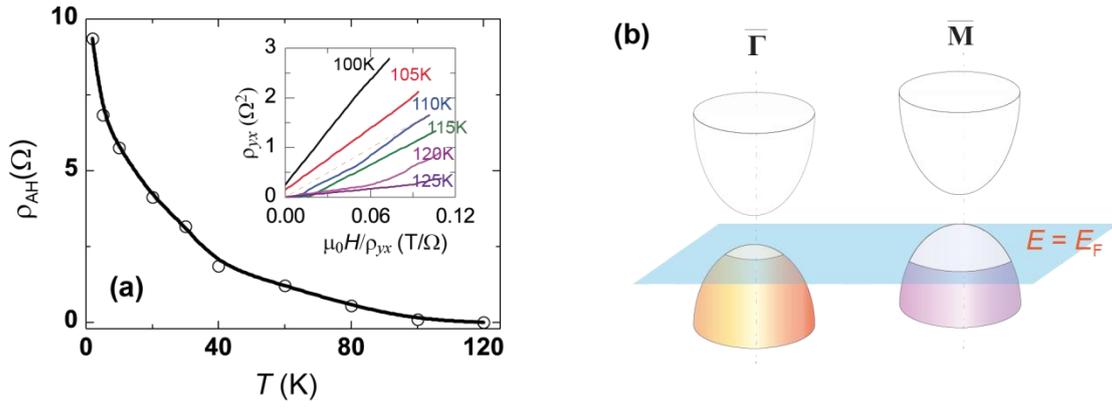

**Fig. 5| Anomalous Hall resistance of 12DL Cr-doped SnTe (111) film.** (a) Temperature dependence of the anomalous Hall resistance ($\rho_{AH}$) of 12DL $Sn_{0.9}Cr_{0.1}Te$ (111) film. Inset: the Arrott plot of 12DL $Sn_{0.9}Cr_{0.1}Te$ (111) film, showing its $T_C$ is ~110K. (b) Schematics of the gapped surface states at $\bar{\Gamma}$ and $\bar{M}$ points in Cr-doped SnTe (111) films when the chemical potential is near the charge neutral point. The binding energy difference of the Zeeman gaps at $\bar{\Gamma}$ and $\bar{M}$ might be reduced by tuning thickness and/or composition of the films, and the chemical potential can be tuned into all four Zeeman gaps.